\documentclass[12pt]{article}
\usepackage[margin=1in]{geometry}
\usepackage{appendix}
\usepackage{amsmath}
\usepackage{graphicx}
\usepackage{lineno}
\usepackage{array}
\usepackage{longtable}
\usepackage{natbib}
\usepackage{amssymb}
\usepackage{geometry}
\usepackage[separate-uncertainty=true]{siunitx}
\usepackage{tabularx}
\usepackage{hyperref}
\usepackage{bm}
\usepackage{fancyvrb}
\usepackage{caption}
\usepackage{booktabs}
\usepackage{makecell}
\usepackage{placeins}
\usepackage{refcount}
\usepackage{lscape}
\usepackage{authblk}

\newcommand{\lwp}{\mathrm{LWP}}
\newcommand{\rref}{\rho_\mathrm{ref}}
\newcommand{\pref}{p_\mathrm{ref}}

\newcommand{\aver}[1]{ \! \langle {#1} \rangle \!}

\begin{document}

\title{Effect of grid anisotropy, resolution, and subgrid-scale models in pseudo-spectral Large Eddy Simulations of low-level clouds}


\author{Davide Selvatici, Richard J.A.M. Stevens}

\affil{Physics of Fluids Group, Max Planck Center Twente for Complex Fluid Dynamics, J. M. Burgers Center for Fluid Dynamics, University of Twente, P.O. Box 217, Enschede, 7500 AE, The Netherlands}

\date{}

\maketitle

\begin{abstract}
We investigate the effect due to grid resolution and subgrid-scale model on large-eddy simulations of low-level clouds using a novel framework that combines pseudo-spectral advection with the anisotropic minimum dissipation (AMD) subgrid-scale model. We use two field campaigns as reference, DYCOMS-II RF01 and ASTEX, which cover both non-precipitating and precipitating stratocumulus cloud regimes across different time scales. Our results demonstrate that the AMD model combined with pseudo-spectral advection produces robust and accurate predictions across varying grid resolutions without parameter tuning. We identify a recommended grid anisotropy where vertical spacing is approximately three times finer than horizontal spacing, balancing accuracy and computational efficiency. Finally, an error analysis based on cloud liquid water content and vertical velocity variance reveals good agreement with theoretical predictions for isotropic grids, while grid anisotropy effectively improves convergence rates.
\end{abstract}

\section{Introduction} \label{sec:intro}
Low-level clouds are the most common type of clouds by annual coverage, with an average of $35\%$ over the oceans and of $17\%$ over the land \citep{lohmann16, hahn07}. These clouds lie within the atmospheric boundary layer (ABL) and are a vital component of the Earth's climate system, as their high albedo reflects a substantial portion of incoming solar radiation, and their infrared opacity generates a cooling at the cloud-top able to strengthen convective mixing in the ABL \citep{wood12,lohmann16,mellado17}. Large Eddy Simulations (LES) have become a valuable tool to study these clouds, as they can resolve key physical processes with greater accuracy than meso-scale models and provide more details than standard measurements.

Despite their high computational cost, stemming from stringent resolution requirements and complex model formulations, LES have been used in cloud research for several decades \citep{breth99, sommeria76, moeng86, cuijpers93}. For example, \cite{breth99} compared various LES codes in simulations of radiatively active smoke layers as analogs for cloud layers. Their study showed that LES can accurately capture the dynamics of radiatively driven entrainment, provided that the vertical grid spacing is fine enough to resolve small-scale features in the cloud-top region. Building on this foundation, \cite{stevens05} evaluated several LES codes using observational data from the DYCOMS-II RF01 campaign \citep{stevens03b}, representing the dynamics of a realistic stratocumulus-topped boundary layer (STBL) and demonstrating that significant differences between codes can arise. This variability is caused by the tendency of the available subgrid-scale (SGS) models to overestimate numerical dissipation, leading to artificial cloud evaporation, and to the use of low-order advection schemes, a choice constrained by the computational overhead of higher-order schemes \citep{mellado18}. 

More recently, advancements in computational resources have facilitated higher resolutions and extended integration times. \cite{mellado18} and \cite{mellado25} performed direct numerical simulations (DNS), achieving good agreement with observations and demonstrating that the achieved Reynolds numbers have reached a level where DNS can effectively reproduce atmospheric states. Moreover, \cite{dussen13} conducted a multi-model comparison for the Atlantic Stratocumulus Transition Experiment (ASTEX) campaign \citep{albrecht95, dussen15-thesis}, which covers a 40-hour diurnal cycle featuring precipitation and a transition from stratocumulus to cumulus clouds. Although satisfactory agreement with observations was obtained, inconsistencies were observed during the precipitation period of this campaign, underscoring the importance of further method development. While the comparison between different microphysical parameterizations lies outside the scope of this study, precipitation models utilize the same numerical schemes and, as such, are susceptible to analogous sensitivities \citep{lamaakel23}.

Developments in the SGS models and numerical advection scheme can have a beneficial effect in enhancing the accuracy of stratocumulus clouds simulations. \cite{matheou19} demonstrated that LES converges toward observational data using isotropic grids when the SGS model implemented by \cite{matheou14} is used. \cite{efsta23} compared constant-coefficient and dynamic SGS models, showing that the former significantly underestimates two key features of STBLs, namely convective mixing and liquid water content. \cite{matheou19} further illustrate that in flow regions where SGS models are less effective, dissipative advection schemes can amplify numerical oscillations, and artificially impact key physical processes such as the cloud-top entrainment velocity. Moreover, \cite{matheou16} study the effect of different advection schemes in an idealized LES setting applied to a sharp one-dimensional pulse, highlighting the superiority of spectral advection. Yet, its use in STBL studies has remained limited \citep{moeng86, stevens05}, and in the intercomparison by \cite{stevens05} it did not outperform the other participating codes.

In this work, we introduce a LES framework featuring pseudo-spectral advection, i.e., spectral accuracy in the horizontal directions and centered finite differences in the vertical, as well as the anisotropic minimum dissipation (AMD) SGS model \citep{rozema15}, applied for the first time to moist boundary-layer simulations. We systematically evaluate the impact of grid resolution, grid anisotropy, and SGS model on the LES accuracy and computational efficiency. In this work, we use DYCOMS-II RF01 and ASTEX as benchmarks to test our framework under both non-precipitating and precipitating cloud conditions. The results demonstrate robust performance even on relatively coarse meshes, without the need for parameter tuning.

The remainder of the paper is organized as follows. Section~\ref{sec:numframe} presents the governing equations and numerical framework characteristics; Section~\ref{sec:res} analyzes the effect of the different numerical choices using the DYCOMS-II RF01 and ASTEX measurement campaigns, comparing the error convergence rate with results available in the literature. Finally, in Section~\ref{sec:conclusions} some concluding remarks are discussed. Technical details of the simulations and initial conditions are reported in the Appendices.

\section{Numerical Framework} \label{sec:numframe}
We extend an in-house LES code previously utilized for both atmospheric and wind energy studies \citep[see, e.g.,][]{ste14d,gad21-blm,stieren22-PRX,kasper24} with state-of-the-art anelastic equations for cloud and precipitation physics. The code uses a pseudo-spectral approach, using spectral derivatives in the horizontal directions and centered, second-order finite differencing in the vertical direction to account for non-periodic boundary conditions in the bottom and top boundaries and allowing grid refinement to be used \citep{moeng84, albertson99, porte00}. The 3/2 anti-aliasing rule mitigates aliasing errors arising from the nonlinear terms producing a non-dissipative advection scheme \citep{canuto12}. Parallelization is implemented in the vertical direction using MPI, and the vertical velocity is defined on a staggered grid. Time integration is performed using a third-order Adams--Bashforth. All simulations employ a CFL number of 0.1, ensuring stability for explicit time integration and following the time step recommendations by \cite{yama12}.

\subsection{Governing equations}
\label{sec:gov_eq}
We adopt the anelastic form of the Navier--Stokes equations to filter out sound waves while retaining essential compressibility effects \citep{bannon96, mellado17}. The corresponding mass and momentum conservation equations are \citep{vallis17, bannon96}:
\begin{equation}
\label{eq:flow}
 \begin{cases}
 \displaystyle \nabla \cdot (\rref \tilde{u}) = 0, \\

 \displaystyle \partial_t \tilde{u} + (\tilde{u} \cdot \nabla) \tilde{u} = - 1/\rref \nabla \tilde{p} + \nabla \cdot \tilde{\tau} + f_c (\tilde{u} - U_G) + g \beta.
 \end{cases}
\end{equation}
Here $\tilde{\cdot}$ represents the filtering operator, $\rref = \rref(z)$ is the background density profile (described in Section~\ref{sec:rref}), $\tilde{u}$ is the filtered velocity field, $\tilde{p}$ is the filtered pressure, $\tilde{\tau}$ is the deviatoric part of the Reynolds stress tensor, $f_c$ is the Coriolis frequency, $U_G$ is the geostrophic velocity above in the free-troposphere, $g = 9.81$ $m/s^2$ is the acceleration due to gravity and $\beta$ is the buoyancy parameter, defined as \citep{vallis17}:
\begin{equation}
 \label{eq:buoy}
 \beta = \frac{\rref - \rho}{\rref}
\end{equation}
where $\rho$ is the local thermodynamic density. The difference between these and the incompressible equations is the presence of $\rref = \rref(z)$ in the continuity equation, which requires a modified pressure Poisson equation in order to find the pressure term $\tilde{p}$ \citep{mellado10a}.

Moisture evolution is modeled using two scalar fields, i.e., the total water humidity, $q_t$, and the liquid water static energy, $h$. Both influence the local density, $\rho$, and thereby contribute to the buoyancy force in the vertical momentum equation (see Section~\ref{sec:gov_eq}). The total water humidity is the sum of vapor and liquid water contents, $q_t = q_v + q_l $; $h$ is the sum of enthalpy and potential energy in the system \citep{mellado18, mellado10a}, i.e., $h = \left [ c_d + q_t (c_v-c_d) \right ] T - q_l L_v + gz$, in which $c_d$ is the specific heat capacity at constant pressure of dry air, $c_v$ for water vapor and $c_l$ for liquid water, $T$ denotes the temperature and $L_v$ is the enthalpy of vaporization. The two scalars are coupled to the flow equations via the buoyancy term $\beta$ (Equation~\eqref{eq:buoy}), and their evolution equations are expressed as 
\begin{equation}
\label{eq:scal}
 \begin{cases}
 \displaystyle \partial_t \tilde{h} + (\tilde{u} \cdot \nabla) \tilde{h} = \nabla \cdot \left (k_h \nabla \tilde{h} - \mathbf{R} + \mathbf{F} \right ) + \omega \partial_z \tilde{h} - \phi_{h}(q_r,N_r), \\

 \displaystyle \partial_t \tilde{q_t} + (\tilde{u} \cdot \nabla) \tilde{q_t} = \nabla \cdot (k_{q_t} \nabla \tilde{q_t}) + \omega \partial_z \tilde{q_t} - \phi_{q_t}(q_r,N_r).
 \end{cases}
\end{equation}
Here $k_h$ and $k_{q_t}$ denote the diffusivity computed from the SGS model (see Section~\ref{sec:sgs}); $\mathbf{R}$ and $\mathbf{F}$ represent the longwave and shortwave radiative forcing terms, respectively (Section~\ref{sec:rad}); and $\omega$ is the prescribed large-scale subsidence velocity. Finally, the effect of precipitation on these scalars is considered through the forcings $\phi_h$ and $\phi_{q_t}$ \citep{heus10}.

To model precipitation, we employ the two-moment bulk microphysics scheme developed by \cite{kk00}, conceived for drizzling stratocumulus clouds \citep{dussen13, kogan03, stevens08}. This scheme introduces two additional scalars, i.e. the rain water mixing ratio, \( q_r \), and the rain droplet number concentration, \( N_r \). Their evolution is governed by the following equations:
\begin{equation}
\label{eq:rain}
 \begin{cases}
 \displaystyle \partial_t \tilde{q_r} + (\tilde{u} \cdot \nabla) \tilde{q_r} - \nabla \cdot (k_{\tilde{q_r}} \nabla \tilde{q_r})= \frac{\partial \upsilon_{\tilde{q_r}} \tilde{q_r}}{\partial z} + \chi^\mathrm{au}_{\tilde{q_r}} + \chi^\mathrm{ac}_{\tilde{q_r}} + \chi^\mathrm{ev}_{\tilde{q_r}}, \\

 \displaystyle \partial_t \tilde{N_r} + (\tilde{u} \cdot \nabla) \tilde{N_r} - \nabla \cdot (k_{\tilde{N_r}} \nabla \tilde{N_r})= \frac{\partial \upsilon_{\tilde{N_r}} \tilde{N_r}}{\partial z} + \chi^\mathrm{au}_{\tilde{N_r}} + \chi^\mathrm{ev}_{\tilde{N_r}}.
 \end{cases}
\end{equation}
%
In Equation~\eqref{eq:rain}, the advection and time integration schemes employed in the rest of the code are used, while the forcing terms contain source terms that depend on $h$ and $q_t$ \citep{heus10}. The term $\partial (\upsilon_{\phi} \phi) / \partial z$ represents the sedimentation of rainwater, where $\upsilon_{\phi}$ is the sedimentation velocity and $\phi \in \{\tilde{q_r}, \tilde{N_r}\}$. The term $\chi^{\mathrm{au}}_{\phi} $ accounts for autoconversion of cloud droplets, while \( \chi^{\mathrm{ac}}_{\tilde{q_r}} \) represents their accretion. The term $ \chi^{\mathrm{ev}}_{\phi}$ describes the evaporation of liquid droplets \citep{kk00, stevens08}.

\subsection{Thermodynamic reference profiles}
\label{sec:rref}
The anelastic equations rely on the reference profiles of pressure, temperature, and density. These profiles are time-invariant and depend on the surface pressure $p_s$ and on the initial profiles of $h$ and $q_t$. Specifically, using these initial conditions, the hydrostatic balance can be solved. It reads as \citep{mellado10a}:
\begin{equation}
 \label{eq:balance}
 \frac{R_d T_0}{g L_0} \frac{d\pref}{dz} + \frac{1}{R_t T}\pref =0, ~~~~~~~ \pref(0)=p_s,
\end{equation}
where $L_0$ denotes the reference length used to nondimensionalize the governing equations. The equation can be integrated vertically to obtain the thermodynamic reference pressure $\pref = \pref(z)$, from which $T_\mathrm{ref}$ and $\rref$ can also be obtained \cite{mellado18,ooyama01,satoh03}. In Equation~\eqref{eq:balance}, $R_t$ is the gas constant of moist air, which depends on $q_l$ and $q_t$, and is defined as \citep{mellado17}:
\begin{equation}
 \label{eq:R}
 R_t = (1-q_t) R_d + q_v R_v = R_d (1 + q_t \varepsilon - q_l (\varepsilon+1)),
\end{equation}
in which $R_d$ is the gas constant of dry air, $R_v$ is the gas constant of water vapor, and $\varepsilon = R_v / R_d - 1 \approx 0.608$. 

\subsection{Condensation and evaporation}
\label{sec:cond}
To determine the equilibrium composition of liquid water humidity, we use the eighth-order polynomial approximation of the Clausius--Clapeyron equation to compute the saturation pressure $p_\mathrm{sat}(T)$ and its temperature derivative, from which $q_l=q_t-q_\mathrm{sat}$ can be found (see \cite{mellado10a, flatau92}).

The methodology is based on the following steps. A provisional temperature is first estimated by assuming all liquid water is in vapor form, and a first estimate of $q_l$ is obtained. If $q_t \leq p_\mathrm{sat}(T)/(\rref R_v T)$, i.e., if the vapor content is below saturation, equilibrium is already satisfied. In this case, we apply a smoothing function to ensure continuity near the saturation threshold, where $q_t = q_\mathrm{sat}$ (e.g., at the cloud base) \citep{mellado10a}. When $q_t > q_\mathrm{sat}$, the local values of $h$ and $q_t$ are used to determine the temperature that satisfies local saturation equilibrium iteratively using the saturation polynomials given by \cite{flatau92, satoh03}. Following \cite{mellado10a}, we solve this problem using a Newton--Raphson method to find the vapor content at equilibrium, $q_v$, or equivalently the liquid water content: $q_l = q_t - q_v$.

%

\subsection{Longwave and shortwave radiative forcing}
\label{sec:rad}
The longwave radiative forcing term, $\mathbf{R}$, describes the net longwave radiation in the atmosphere. As shown by \cite{larson07}, it can be approximated using this expression \citep{mellado17,stevens05}:
\begin{equation}
\label{eq:rad}
 R(z) = F_0 e^{\left ( -k  \lwp (z) \right )} - F_1 e^{\left ( -k \left (\lwp(0) - \lwp(z)\right ) \right )},
\end{equation}
in which $\lwp(z)$ is the liquid water path at the height $z$, defined as \cite{stevens05,wood12,mellado17}:
\begin{equation}
 \label{eq:lwp}
 \lwp(z) = \int_z^{\infty} \rref  q_l  d{z'}.
\end{equation}
$\lwp(z)$ represents the amount of liquid water for every vertical column, and is computed for every grid point at each time step. In the following, with $\lwp$ we refer to the total column-integrated water content, i.e., starting from $z=0$. In Equation~\eqref{eq:rad}, $F_0$, $F_1$, and $k$ are constant parameters obtained by fitting measurements-based estimates, single-column or general circulation models \citep{stevens05, ackerman09, fu93}.

Unlike most atmospheric codes, which typically parallelize in the horizontal directions, the numerical framework used in this work relies on vertical parallelization for scalability due to the pseudo-spectral advection. To avoid the computational overhead associated with the vertical integration required to compute $\lwp$, this operation is performed using a dedicated MPI routine, \verb|MPI_Scan|, that maximizes efficiency.

To model shortwave radiative heating $\mathbf{F}$, we use the delta-Eddington approximation \citep{briegleb92, joseph76}, which accounts for scattering and absorption based on the effective radius $r_\mathrm{eff}$, i.e., a representative measure of the cloud droplet size distribution \citep{steph84,dussen13, breng00}.
%
This is used to find the cloud optical depth $\tau$ as
\begin{equation}
 \label{eq:tau}
 \tau(z) = \frac{3}{2} \frac{\lwp(z)}{r_\mathrm{eff} \rho_l},
\end{equation}
which is then used to find the shortwave radiative flux:
\begin{equation}
 F(z) = \frac 43 S_0 \left ( p \left ( c_1 \exp (-k \tau') - c_2 \exp (k \tau') \right ) - \beta \exp (-\tau'/\mu) \right ) + \mu S_0 \exp (-\tau'/\mu),
\end{equation}
in which $S_0$ is the constant for incoming shortwave radiation at the cloud-top; $\mu$ is the cosine of the solar zenith angle; $c_1$ and $c_2$ are two variables which depend on the local value of $\tau$ and the prescribed surface albedo; $\tau'$ is defined as $\tau' = (1-\omega_e g_a^2) \tau$, where $\omega_e$ is the equivalent single scattering albedo, and $g_a=0.85$ is the asymmetry factor of droplet scattering distribution \cite{fouquart89, clatchey71}; $p$, $k$, and $\beta$ depend on the local $\tau$, $g_a$ and $\mu$. For more details about the radiative heating model, we refer to \cite{heus10} and \cite{ackerman09}.

\subsection{Anisotropic Minimum Dissipation SGS model for moist boundary layers}
\label{sec:sgs}
In Equations \eqref{eq:flow}, \eqref{eq:scal} and \eqref{eq:rain}, the diffusion terms are modeled using the AMD model introduced by \cite{rozema15, rozema20}. Here we extend the AMD model to account for moisture and precipitation scalars. The stress tensor and the associated fluxes of heat and moisture are represented as follows:
\begin{align}
\begin{split}
 \label{eq:sgs1}
 \tau_{ij} - \frac 13 \tau_{kk} &= - 2 \nu_t \tilde{S}_{ij}, \\
 k_\Phi \partial_j \tilde{\Phi} &= - \nu_\Phi \partial_j \tilde{\Phi},
\end{split}
\end{align}
where $\Phi = \{ h, q_t, q_r, N_r \}$ and $\tilde{S}_{ij} = \frac 12 (\partial_j \tilde{u}_i + \partial_i \tilde{u}_j)$ is the filtered strain rate tensor and $|\tilde{S}| = \sqrt{2 \tilde{S}_{ij} \tilde{S}_{ij}}$. The trace of the SGS stress tensor is absorbed into the filtered modified pressure $\tilde{p}^* = \tilde{p} - p_\infty + \mathrm{Tr}(\bm{\tau})/3$, where $\tilde{p}$ is the kinematic pressure. 


The AMD model enforces an upper bound on the sub-filter-scale energy within a box $\Omega_b$ of dimensions $\Delta_x \times \Delta_y \times \Delta_z$:
\begin{equation}
 \label{eq:poincare}
 \int_{\Omega_b} \frac 12 \tilde{u}'_i \tilde{u}'_i dV \leq C \int_{\Omega_b} \frac 12 (\Delta_i \partial_i \tilde{u}_j) (\Delta_i \partial_i \tilde{u}_j) dV.
\end{equation}
Here $C$ is a constant which depends on the advection scheme adopted. Following \cite{rozema20,abkar16}, we set $C = 1/12$ in the horizontal directions, where a spectral method is used, and $C = 1/3$ in the vertical directions, where we employ centered finite differences. By applying the procedure as in Equation~\eqref{eq:poincare} to $h$ and $q_t$, and using the midpoint rule to evaluate the integral, the SGS viscosity is obtained.

The resulting expression for the diffusivity for the moisture scalars is the following:
\begin{equation}
 \label{eq:nuscal}
 \nu_\Phi = C \frac{(\Delta_k \partial_k \tilde{u}_i) (\Delta_k \partial_k \tilde{\Phi}) \partial_i \tilde{\Phi}}{\partial_l \tilde{\Phi} \partial_l \tilde{\Phi}}.
\end{equation}

The diffusivity for the momentum equation incorporates the following buoyancy correction proposed by \cite{abkar17}:
\begin{equation}
 \label{eq:nut}
 \nu_t = C \frac{- (\Delta_k \partial_k \tilde{u}_i)(\Delta_k \partial_k \tilde{u}_j) \tilde{S}_{ij} + \delta_{i3} (\Delta_k \partial_k \tilde{u}_i) (\Delta_k\partial_k \beta)}{\partial_l \tilde{u}_m \partial_l \tilde{u}_m}.
\end{equation}
%

\subsection{Boundary conditions at the ground}
At the first grid level above the ground we impose a wall-stress boundary condition based on the logarithmic law, using test-filtered velocities to ensure that the mean predicted stress aligns with that given by the logarithmic law \citep{ste14d, moeng84, bouzeid05}. The resolved velocities and scalar values at the first grid point are used to determine the surface shear stress:
\begin{equation}
\label{eq:tau_3}
 \tau_{i3|w}=-u_*^2 \frac{\tilde{u}_i}{\tilde{u}_\mathrm{rel}} = - \left ( \frac{\tilde{u}_\mathrm{rel} \kappa}{\ln(z/z_0) - \psi_M} \right )^2 \frac{\tilde{u}_i}{\tilde{u}_\mathrm{rel}}
\end{equation}
in which $u_*$ is the friction velocity, $\kappa = 0.4$ is the von K\'arm\'an constant, $\tilde{u}_\mathrm{rel} = \sqrt{\tilde{u}^2+\tilde{v}^2}$ is the filtered velocity magnitude at the first grid level, and $z_0$ is the roughness length.

The boundary conditions for the vertical gradients of $h$, $q_t$ are computed using Monin--Obukhov similarity theory \citep{moeng84}. The heat and moisture fluxes are
\begin{equation}
\label{eq:q_scal}
 q_h = \frac{ u_* \kappa (h_s - \tilde{h}) }{\ln \left ( \frac 12 \Delta z / z_0 \right ) - \psi_H}, ~~~~~~~ \; q_{q_t} = \frac{ u_* \kappa (q_s - \tilde{q_t}) }{\ln \left ( \frac 12 \Delta z / z_0 \right ) - \psi_H}.
\end{equation}
where $\tilde{h}$ and $\tilde{q_t}$ are the filtered scalars at the first grid level above the ground. The term $q_s$ is the saturation water humidity, based on the imposed sea-surface temperature, and $h_s$ is computed based on $q_s$ and the sea-surface temperature. The values of $\psi_M$ (Equation~\eqref{eq:tau_3}) and $\psi_H$ (Equation~\eqref{eq:q_scal}) are determined based on the stability of the boundary layer, which depends on the Obukhov length $L= - u_*^3 / \left (\kappa g \phi_\beta \right )$ with $\phi_\beta$ is the surface buoyancy flux \citep{cuijpers93}.

\section{Results}
\label{sec:res}
In the following, our numerical framework is used to simulate the DYCOMS-II RF01 (Section~\ref{sec:dycoms}) and ASTEX (Section~\ref{sec:astex}) cases. These two complex test cases are used here to systematically investigate which numerical factors contribute most significantly to improving computational efficiency and improving convergence rates focusing primarily on the effects of grid resolution, grid anisotropy, and SGS model. As the DYCOMS-II RF01 campaign considers a 4-hour simulation period, it allows us to perform higher-resolution simulations, which we use to obtain detailed information about an optimal grid aspect ratio, error convergence rates, and the efficiency gains of grid anisotropy. The second case, ASTEX, represents a more complex scenario, covering a 40-hour simulation of a stratocumulus-to-cumulus transition through a complete diurnal cycle, including also precipitation. The latter scenario is used to assess the robustness of the results of the numerical framework across different grid discretizations and two SGS models.

\subsection{\textit{DYCOMS-II RF01} case}
\label{sec:dycoms}
\begin{figure}[t]
\centering
\includegraphics[width=0.85\textwidth]{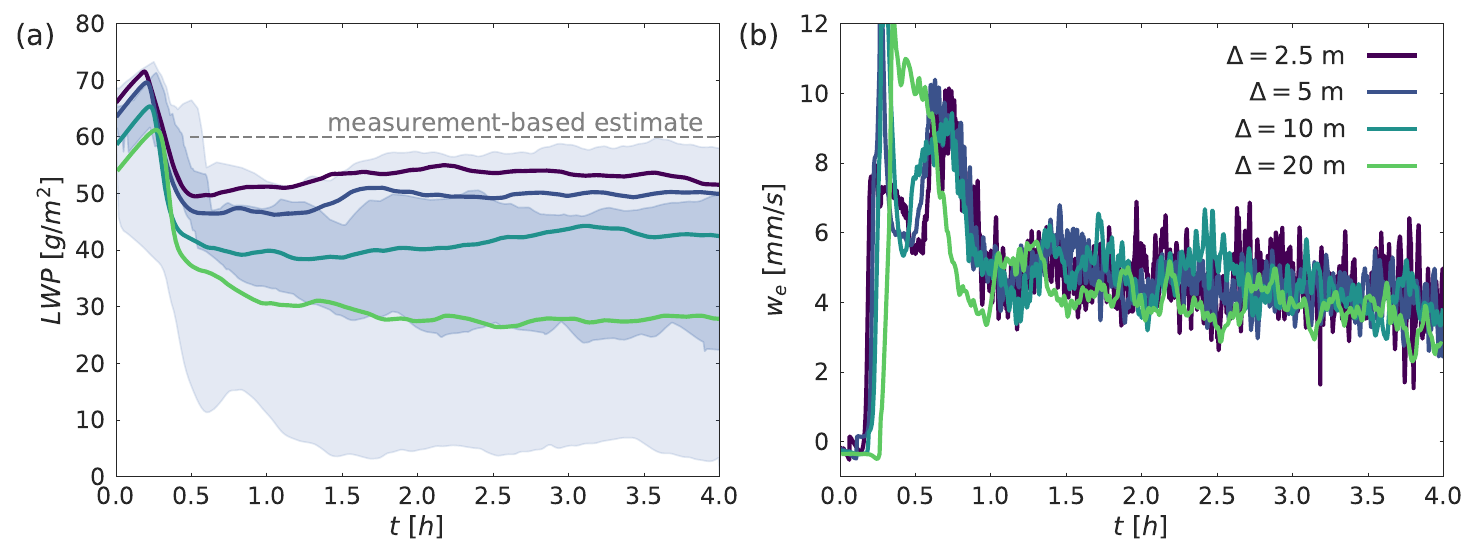}
\includegraphics[width=\textwidth]{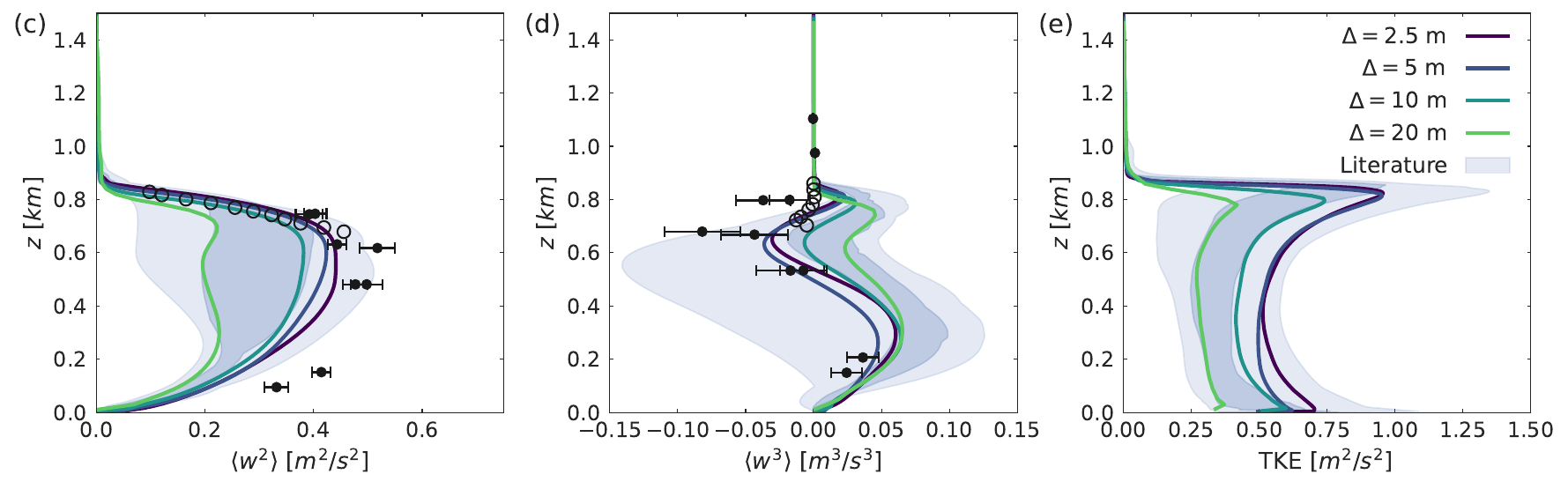}
\caption{Comparison of isotropic grids at different resolutions for the DYCOMS-II RF01 case. (a): $\lwp$, (b): entrainment velocity, (c): vertical velocity variance, (d): third moment of the vertical velocity, (e): turbulent kinetic energy. Lines represent simulation results averaged over the fourth hour; markers indicate measurements as derived in situ (black dot) and radar (circles), taken from \cite{stevens05}. Light shadings delimit maximum and minimum values within the LES codes comparison by \cite{stevens05}; dark shading denotes the central half of the distribution as delimited by the first and third quartiles.}
\label{fig:res_lwp_vel}
\end{figure}
\begin{figure}[t]
\centering
\includegraphics[width=\textwidth]{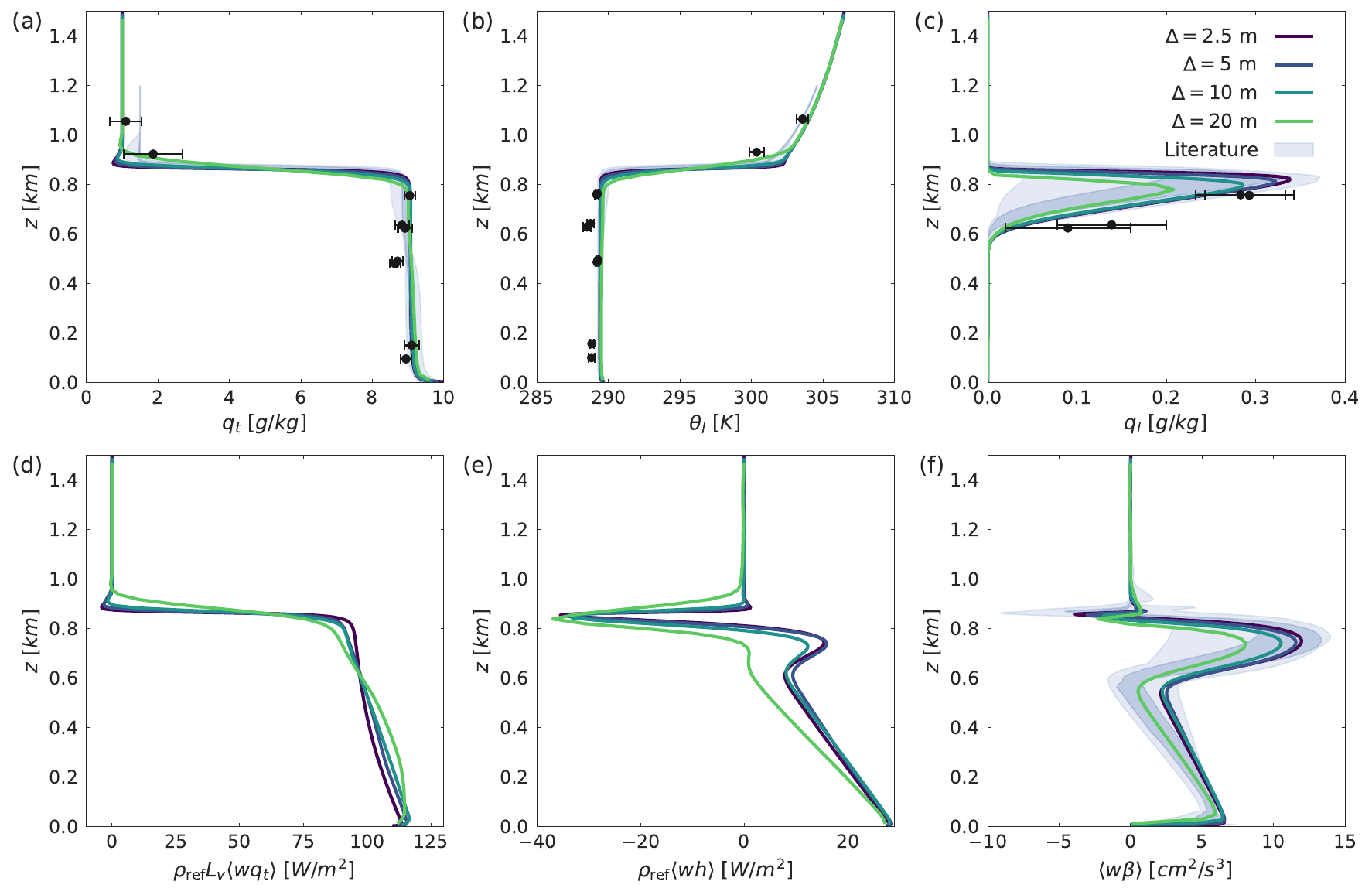}
\caption{Comparison of isotropic grids at different resolutions for the DYCOMS-II RF01 case. (a): total water humidity, (b): liquid water potential temperature, (c): liquid water humidity, (d): total latent heat flux, (e): total liquid water energy flux, (f): resolved buoyancy flux. Results are averaged over the fourth hour. Refer to the caption of Figure~\ref{fig:res_lwp_vel} for details on measurements, markers, and shadings.}
\label{fig:res_anel}
\end{figure}

The initial conditions adopted for the DYCOMS-II RF01 case are reported in Appendix~\ref{app:dycoms} and further information on the measurements and previous simulation attempts can be found in the works by \cite{stevens03b, stevens05, matheou19, mellado18}. Following \cite{matheou19}, the domain size is $\SI{5.12}{km} \times \SI{5.12}{km} \times \SI{1.5}{km}$ in the streamwise, lateral and vertical directions, respectively. 

The highest resolution employed is $\SI{2.55e9}{}$ grid points, requiring about $6\cdot 10^6$ CPU hours. In this case, the grid is discretized using $2048 \times 2048 \times 608$ grid points, corresponding to an isotropic grid spacing of $\Delta = \SI{2.5}{m}$. The other simulations with isotropic grid spacing are: $1024 \times 1024 \times 304$ ($\Delta = \SI{10}{m}$), $512 \times 512 \times 152$ ($\Delta = \SI{5}{m}$), and $256 \times 256 \times 76$ ($\Delta = \SI{20}{m}$). Further details of all simulations performed are reported in Appendix~\ref{app:sim_list} along with information about their computational performance.


\subsubsection{Effect of resolution: isotropic grids} \label{sec:isotropic}
Figure~\ref{fig:res_lwp_vel}a,b shows the temporal evolution of two key STBL parameters in simulations employing isotropic grids ($\Delta \in {2.5, 5, 10, 20}$ m). The two properties are the $\lwp$ (Equation~\ref{eq:lwp}) and the entrainment velocity, defined as $w_e = d z_i(t) / dt + D z_i(t)$, in which $z_i(t)$ is the time-dependent cloud-top height, defined as the interpolated height at which the relative humidity reaches $50\%$, and $D$ is the subsidence rate (Appendix~\ref{app:dycoms}) \citep{mellado18, matheou19}.

During an initial transient phase, in which the $\lwp$ peaks due to simulation spin-up effects \citep{stevens05}, it subsequently decreases and approaches a quasi-steady state, whose value is strongly linked to $\Delta$. At the finest grid spacing, $\lwp$ is about \SI{53}{g/m^2} in the fourth hour, closely matching the observation estimate of \SI{60}{g/m^2}. Discrepancies between the initial $\lwp$ in the simulations arise from the differences in the discretization of the liquid water humidity peak on the different grids (see also Section~\ref{app:smooth}), further highlighting the effect of coarser grids.

As shown in panel b, in all our simulations $w_e$ converges to approximately $4-4.1 \ \si{mm/s}$, consistent with the measured entrainment of \SI{3.9 \pm 0.1}{mm/s} by \cite{stevens03b}, even on coarser grids. Although the $\lwp$ shows higher variability among the simulations, both $\lwp$ and $w_e$ are robust to the grid variation, likely due to the non-dissipative advection scheme adopted (see Section \ref{sec:numframe} and \cite{canuto12}). Supporting this hypothesis, \cite{matheou19} show that lower-order and dissipative advection schemes can introduce numerical errors strong enough to affect boundary-layer dynamics and can lead to an overestimation of $w_e$ by up to 50\% using coarse resolutions.

Figure~\ref{fig:res_lwp_vel}c-e illustrates the influence of grid resolution on velocity statistics. Panel (c) shows the vertical velocity variance $\aver{w^2}$. This quantity indicates the strength of convection within the boundary layer, which depends on the two main forcings, i.e., the surface fluxes and cloud-top radiative cooling (Equation~\eqref{eq:rad}) \citep{mellado17}. Therefore, as the latter is directly influenced by $\lwp$, $\aver{w^2}$ shows a similar resolution sensitivity to $\lwp$. A comparable dependence is observed for $\aver{w^3}$, particularly within the cloud layer. The lowest-resolution cases fail to capture the negative skewness linked to downward convective plumes, setting an upper bound on the vertical spacing of $\Delta \sim 5-10$ m. Consistently, the main deviation between latent heat, sensible heat, and buoyancy flux profiles is seen in the same cloud-top region. This shortcoming is mainly attributed to inadequate resolution near the cloud top, which governs such convective motions due to the localized peak of the radiative forcing in this region.

Figure~\ref{fig:res_lwp_vel}e shows the comparison of the turbulent kinetic energy, defined as $\text{TKE} = \frac 1 2 \sum_i^3 \langle u_i'^2 \rangle$, where $\langle u_i'^2 \rangle = \langle u_i^2 \rangle - \langle u_i \rangle^2$. TKE is severely underestimated for the coarsest meshes considered, while it shows signs of convergence at high resolutions, also due to the improved representation of the vertical velocity variance with higher resolution (Figure~\ref{fig:res_lwp_vel}c).

Figures~\ref{fig:res_anel}a and b show the mean profiles of $q_t$ and liquid water potential temperature $\theta_l$, averaged during the fourth simulation hour. Overall, the profiles show modest sensitivity to grid resolution. A slight underestimation of $q_t$ is observed just above the inversion-layer height (panel a), however, this effect is much smaller than in the simulations by \cite{matheou19} in which a lower-order dissipative advection scheme was tested, likely because of the higher spectral accuracy in the horizontal directions used in this work. Finally, consistently with the achieved $\lwp$, the profile of $q_l$ agrees well with the measurements (Figure~\ref{fig:res_anel}c). The latent heat flux (panel~d), total liquid water energy flux (panel~e), and buoyancy flux (panel~f), are consistent with previous works employing higher resolutions \citep{mellado18}.

\subsubsection{Effect of resolution: anisotropic grids} \label{sec:aniso}

\begin{figure}[t]
\centering
\includegraphics[width=\textwidth]{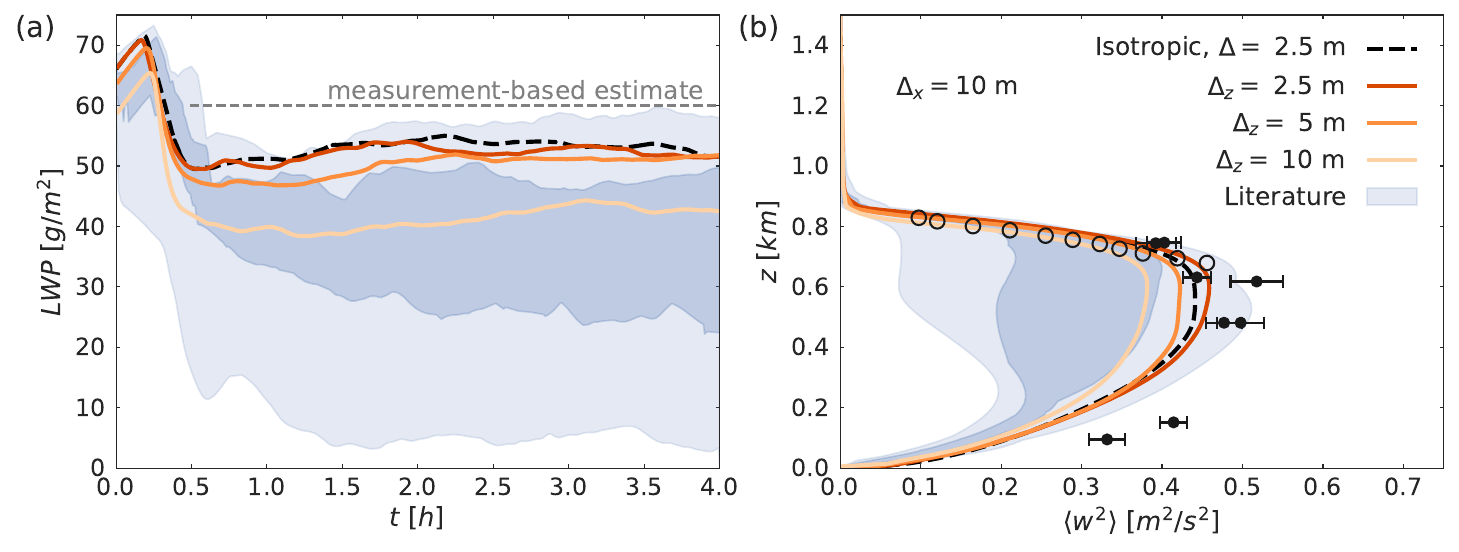}
\includegraphics[width=\textwidth]{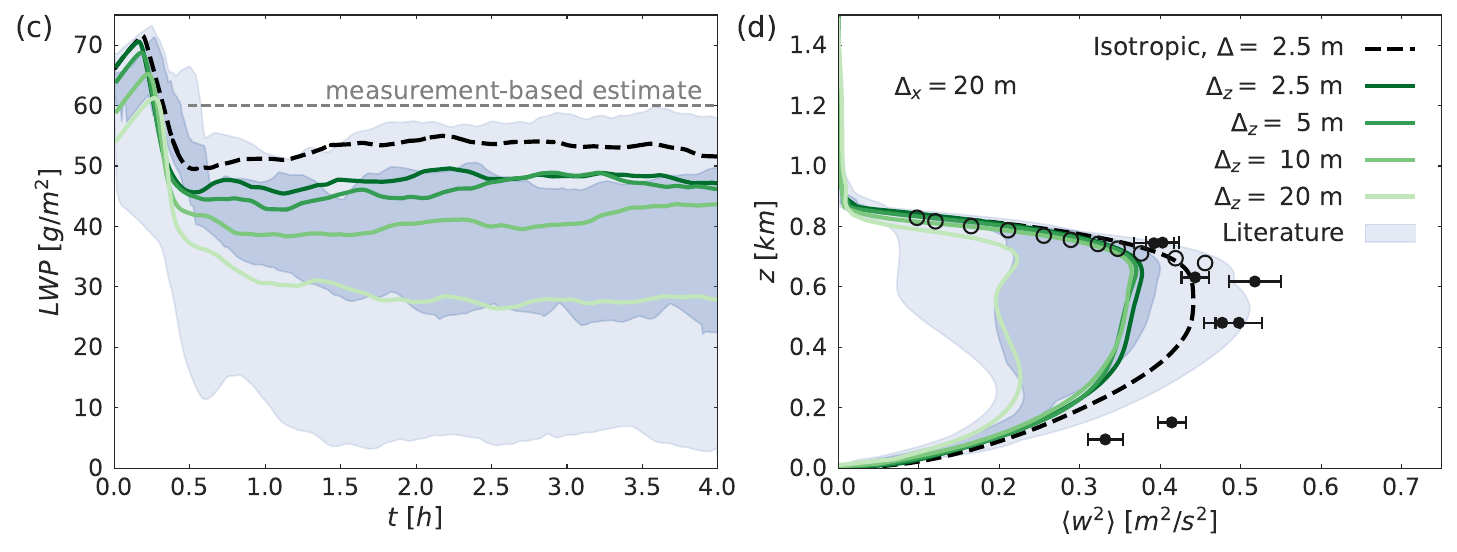}
\caption{Effect of grid anisotropy in the DYCOMS-II RF01 case. Panels (a), (b) use $\Delta_x=\SI{10}{m}$; panels (c), (d) use $\Delta_x=\SI{20}{m}$. (a), (c): $\lwp$; (b), (d): vertical velocity variance. Results are averaged over the fourth hour. Refer to the caption of Figure~\ref{fig:res_lwp_vel} for details on measurements, markers, and shadings.}
\label{fig:aniso}
\end{figure}

The AMD model inherently accounts for grid anisotropy, making it well-suited for simulations with anisotropic grids (see Section~\ref{sec:sgs}). This feature is particularly beneficial for stratocumulus cloud simulations, where enhanced vertical resolution can be effectively used to resolve the small-scale structures in the cloud-top layer, which are characterized by strong gradients of the advected scalars and the radiative forcing \citep{mellado17}.

Figure~\ref{fig:aniso} shows the $\lwp$ and vertical velocity variance obtained for two horizontal spacings, namely $\Delta_x = \SI{10}{m}$ (panels a-b) and $\Delta_x = \SI{20}{m}$ (panels c-d), each tested with vertical grid spacings ranging from \SI{20}{m} to \SI{2.5}{m}. These are compared against the highest-resolution reference case with uniform spacing of $\Delta = \SI{2.5}{m}$. Figure~\ref{fig:aniso} shows that for $\Delta_{x,y} = \SI{20}{m}$, reducing the vertical spacing from $\Delta_z = \SI{20}{m}$ to $\SI{10}{m}$ leads to a substantial improvement in accuracy, with relative errors in the mean $\lwp$ and the vertical velocity variance integral decreasing by more than $50\%$ (see table \ref{tab:lwp_w2} and Section~\ref{sec:rel_error}). Further refinement of $\Delta_z$ yields only marginal benefits. Similarly, for simulations with $\Delta_{x,y} = \SI{10}{m}$, finer vertical resolutions ($\Delta_z = \SI{5}{m}$ and $\SI{2.5}{m}$) improve results compared to $\Delta_z = \SI{10}{m}$, however, the improvement remains modest when $\Delta_z$ is reduced from \SI{5}{m} to \SI{2.5}{m}, with the integral error in $\langle w^2 \rangle$ decreasing by only $3\%$. Therefore, an optimal grid spacing in the vertical direction can be identified in the range $\Delta_z \sim 5-\SI{10}{m}$, beyond which the marginal reduction in error no longer justifies the additional computational cost. Further quantitative error analysis is provided in Section~\ref{sec:rel_error}.

\subsubsection{Effect of the SGS model} \label{sec:smag}

\begin{figure}[t]
\centering
\includegraphics[width=\textwidth]{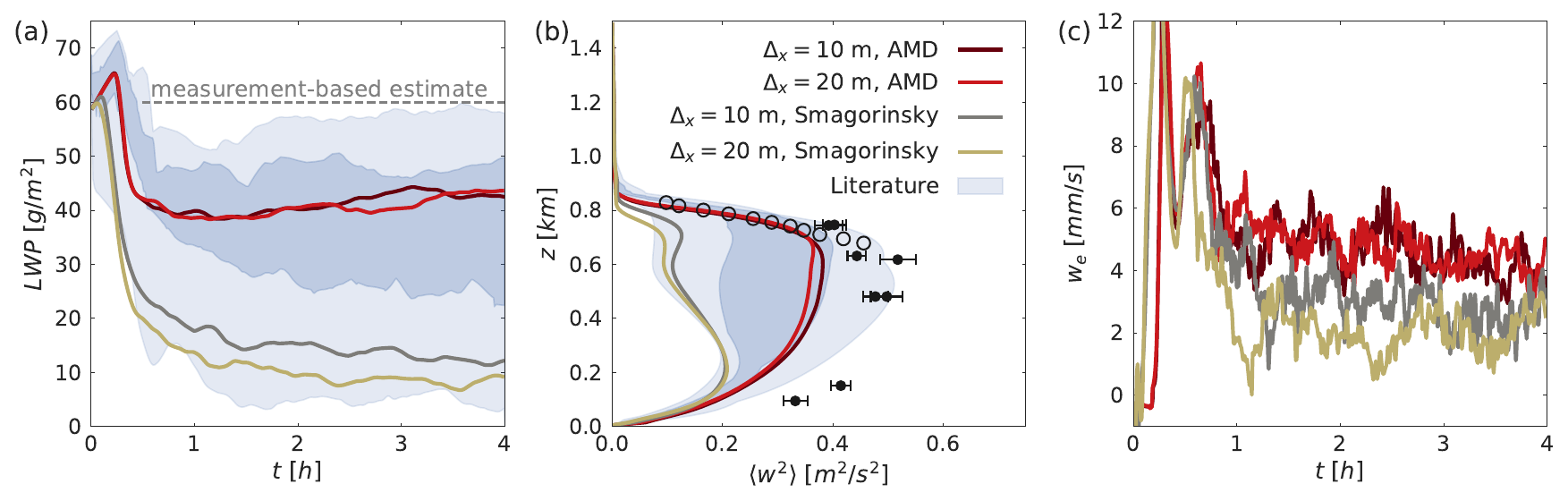}
\caption{Comparison of AMD and Smagorinsky SGS models in the DYCOMS-II RF01 case. All simulations use $\Delta_z = \SI{10}{m}$. (a): $\lwp$ evolution, (b): vertical velocity variance, (c): entrainment velocity. Results are averaged over the fourth hour. Refer to the caption of Figure~\ref{fig:res_lwp_vel} for details on measurements, markers, and shading.}
\label{fig:smag_dycoms}
\end{figure}

In Figure~\ref{fig:smag_dycoms} we compare two SGS models, namely Smagorinsky and AMD, in terms of their effect on $\lwp$, $\langle w^2 \rangle$, and the entrainment velocity, $w_e$. For the Smagorinsky model we use $c_s = 0.18$ and $\mathrm{Pr} = 0.4$ \citep[following][]{stevens05}, while the AMD model does not require any tunable parameters (see also Section~\ref{sec:sgs}). As shown in panel~(a), the simulations using AMD consistently yield $\lwp \approx \SI{43}{g/m^2}$. In contrast, the Smagorinsky model produces $\lwp \approx \SI{10}{g/m^2}$, thus severely underestimating measurements. The underestimation can be attributed to the excessive diffusion in the Smagorinsky model, which overestimates the dissipation of $h$ and $q_t$ in the cloud-top region, where the two scalars have their maximum gradients. This then leads to a dissipation of $q_l$, or equivalently of $\lwp$. The vertical velocity variance (panel~b) is also significantly underestimated in the Smagorinsky simulations, due to the feedback between convection intensity and radiative forcing through the $\lwp$ \citep{kirk06}. As a result, the Smagorinsky model underestimate entrainment at the cloud top (panel~c), yielding entrainment velocities of $w_e \approx \SI{2.5}{mm/s}$ and $w_e \approx \SI{2.0}{mm/s}$ when averaged over the fourth hour. In contrast, the AMD model yields higher entrainment velocities of $w_e \approx \SI{4.1}{mm/s}$ and $w_e \approx \SI{4.3}{mm/s}$ for $\Delta_x = \SI{10}{m}$ and $\Delta_z = \SI{20}{m}$, respectively, which are more consistent with observational estimates. This highlights the importance of the SGS model in accurately representing STBLs dynamics, with the AMD model demonstrating substantially improved performance. While updated coefficients in the Smagorinsky model could improve the accuracy of the results, their determination requires case-specific tuning, which is not necessary when using the AMD model.

\subsubsection{Computational efficiency} \label{sec:rel_error}

\begin{figure}[t]
\centering
\includegraphics[width=\textwidth]{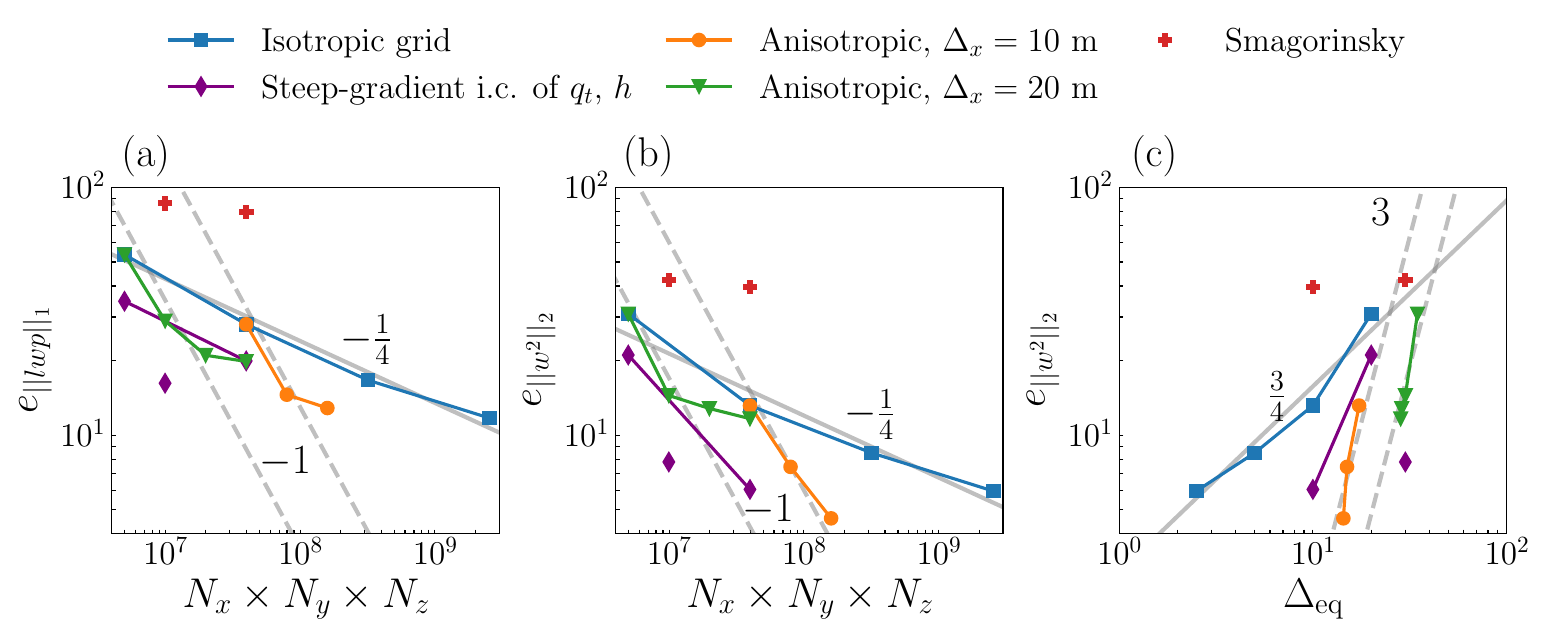}
\caption{Effect of grid anisotropy on numerical convergence for the DYCOMS-II RF01 case. (a): Relative error of $\lwp$, (b): relative error of $\aver{w^2}$, (c): relative error of $\aver{w^2}$ compared to the equivalent grid spacing. Simulations indicated as “Steep gradient i.c. of $q_t$ and $h$" are reported in Appendix \ref{app:smooth}. Formulations of the relative errors are reported in Equation~\eqref{eq:error}. Data points corresponding to simulations with the Smagorinsky SGS model are not connected as they make use of both isotropic and anisotropic grids.}
\label{fig:conv_tot}
\end{figure}
\begin{figure}[t]
\centering
\includegraphics[width=\textwidth]{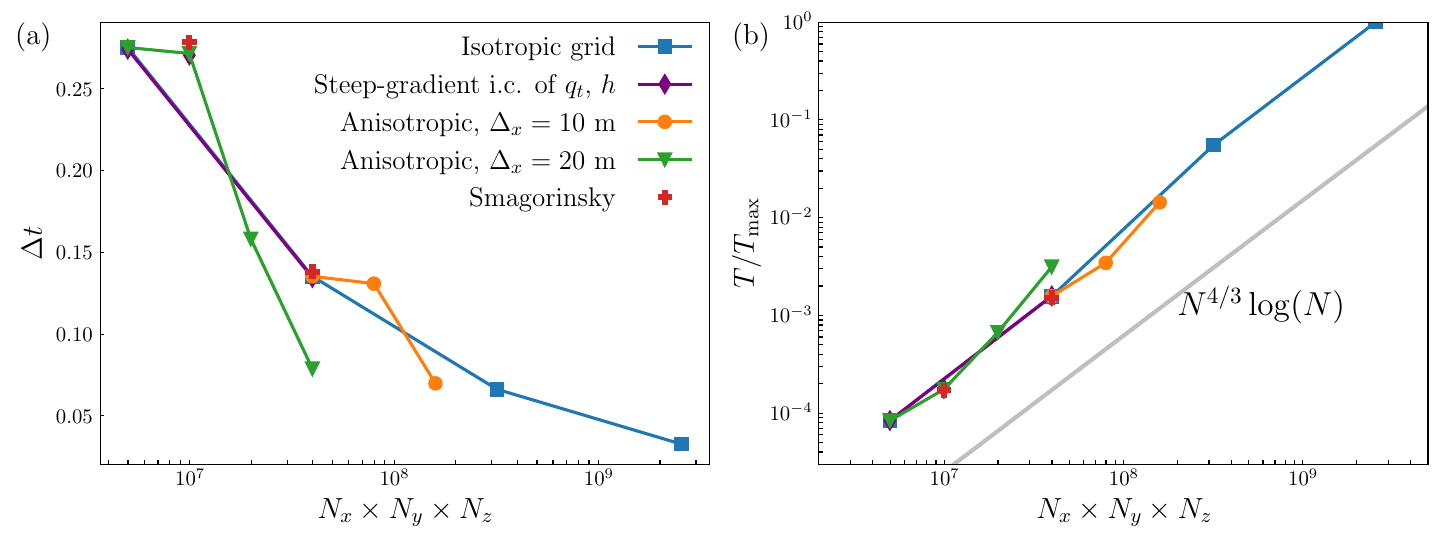}
\caption{Computational aspects of simulations in the DYCOMS-II RF01 case. (a): Average time step in the simulation, (b): normalized computational cost.}
\label{fig:scaling}
\end{figure}
We now estimate the numerical error of the simulations as a function of grid resolution using two metrics: the relative error in the average $\lwp$, and the relative error in the integral of $\aver{w^2}$. As shown before, these quantities are chosen because they capture key STBL dynamics and, at the same time, can be computed directly from the available measurements. We define them as:
\begin{equation}
 e_{||lwp||_1} = \frac{\displaystyle \int_0^{L_z} \rref q_l dz - \lwp_\mathrm{ref}}{\displaystyle \lwp_\mathrm{ref}} \ \ \ \text{and} \ \ \ e_{||w||_2} = \frac{\displaystyle \left ( \int_0^{L_z} \aver{w^2}\ dz \right )^{1/2} - \left ( \int_0^{L_z} \aver{w^2}\ dz \right )_\mathrm{ref} ^{1/2}}{\displaystyle \left ( \int_0^{L_z} \aver{w^2}\ dz \right )_\mathrm{ref} ^{1/2}},
\label{eq:error}
\end{equation}
where $\lwp_\mathrm{ref}$ is the measurement-based estimate of $\SI{60}{g/m^2}$, while $\left (\int \aver{w^2} \right )_\mathrm{ref}^{1/2}$ is estimated from the measurements data (Figure~\ref{fig:res_lwp_vel}c), and it is equal to $\SI{18.06}{m^{3/2}/s}$. Figures~\ref{fig:conv_tot}a,b show that both $e_{||lwp||_1}$ and $e_{||w||_2}$ scale approximately as $N^{-1/4}$ for isotropic grids, where $N$ is the total number of grid points.

\cite{duran19} used LES with AMD and Smagorinsky models to study error scaling in plane channel flows. They show that grid anisotropy can be taken into account in the scaling by considering an equivalent grid spacing, $\Delta_\mathrm{eq} = \sqrt{\Delta_{x}^2 + \Delta_{y}^2 + \Delta_{z}^2}$. They find numerical error scalings for the main turbulent statistics far from the wall, and compare these errors with theoretical estimates. They conclude that the relative error on vertical velocity fluctuations scales as $\Delta_\mathrm{eq}^{0.4-0.8}$, which is consistent with our scaling for isotropic grids (Figure~\ref{fig:conv_tot}c). Specifically, in the case of $e_{||w||_2}$, the scaling is $\sim \Delta_\mathrm{eq}^{3/4}$ (Figure~\ref{fig:conv_tot}c), and in the case of $e_{||lwp||_1}$ a similar exponent is found (not shown). Furthermore, \cite{duran19} show that the relative error in the mean velocity scales approximately as $\epsilon \Delta_\mathrm{eq}$, where $\epsilon$ is a constant that depends on the SGS model. For the AMD model, this constant is lower, indicating higher accuracy and aligning with our findings in Section~\ref{sec:smag}.

Figure~\ref{fig:conv_tot} shows that the error decreases significantly faster when anisotropic grids are used. This can be traced back to the physical characterization of STBLs, in which the primary source of turbulent fluctuations is located in the cloud-top region at the top of the boundary layer (see also \cite{mellado10b,mellado17}). For this reason, anisotropic grids are more effective in reducing the error reduction, whose rate can reach up to $N^{-1}$ or $\Delta_\mathrm{eq}^3$ in the simulations considered. However, as mentioned in Section~\ref{sec:aniso}, this trend features an optimal aspect ratio between 2 and 4, above which the results' accuracy is mainly unaffected and computational costs increase. Finally, the steep gradients in the initial vertical profiles of $q_t$ and $h$ in the cloud-top region (Appendix~\ref{app:smooth}) may accelerate the error reduction rate even more, as Figure~\ref{fig:conv_tot}b,c show.

Each simulation average time step is reported in Figure~\ref{fig:scaling}a and Table~\ref{tab:lwp_w2}. Since the magnitude of the horizontal velocity components is higher than the fluctuations of vertical velocity, at low resolutions or for isotropic grids the time step is not affected by the vertical grid spacing. However, for $\Delta_x=\SI{20}{m}$ the time step of the simulations with $\Delta_z = \SI{5}{m}$ and $\Delta_z = \SI{10}{m}$ differ by a factor of 1.7; similarly, halving $\Delta_z$ from 5 m to 2.5 m cuts the time step further in half. Furthermore, for $\Delta_x=\SI{10}{m}$, $\Delta t$ for simulations with $\Delta_z = \SI{2.5}{m}$ and $\Delta_z = \SI{5}{m}$ differ by approximately a factor of 2, while it is almost constant when $\Delta_z$ is doubled again. Hence, for sufficiently high resolution, the vertical grid spacing becomes more restrictive than the horizontal one, causing higher computational cost. Figure~\ref{fig:scaling}b and Table~\ref{tab:lwp_w2} report the computational time, showing that the cost of each simulation closely follows the expected scaling, which accounts for both the spectral convection contribution, $N \log(N)$, and the time-step contribution, $N^{1/3}$. Only simulations using strongly anisotropic meshes with large aspect ratios deviate from this trend. However, these cases are not optimal from an accuracy-cost perspective.
\begin{figure}[t]
\centering
\includegraphics[width=0.65\textwidth]{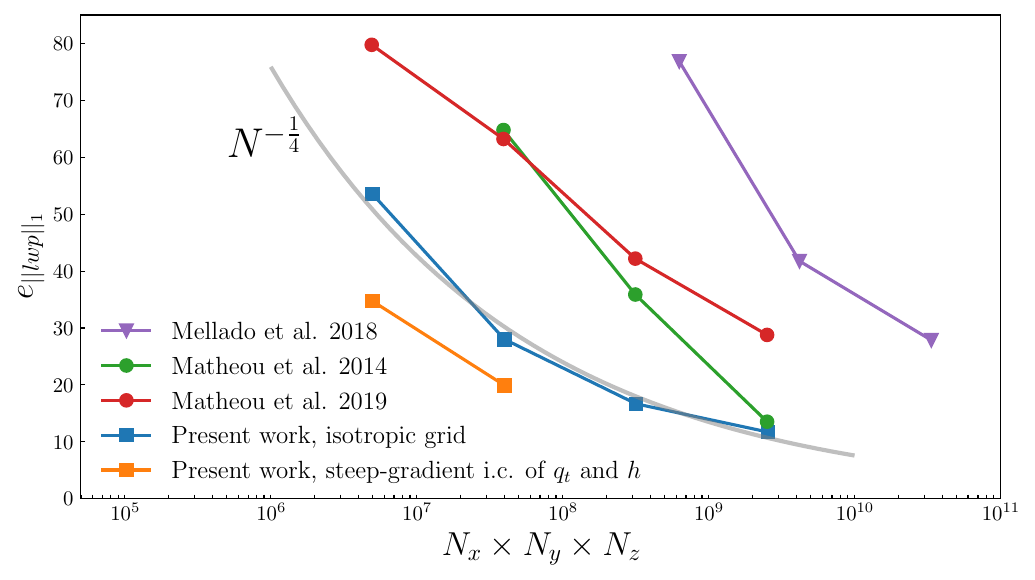}
\caption{Relative error of $\lwp$ (Equation~\ref{eq:error}) for the DYCOMS-II RF01 case. The relative error calculated for the works of \cite{mellado18} and \cite{matheou14, matheou19} are based on digitally extracted data from these references. Simulations indicated as “steep gradient i.c. of $q_t$ and $h$" are reported in Appendix \ref{app:smooth}. Grey line indicates the error scaling of $N^{-1/4}$, $N$ being the grid resolution.}
\label{fig:conv_codes}
\end{figure}

Finally, Figure~\ref{fig:conv_codes} compares the variation of $e_{||lwp||_1}$ between the present study and the results by \cite{matheou14, matheou19} and \cite{mellado18}, which employ different numerical methods. Specifically, \cite{matheou14, matheou19} use a finite-difference-based LES framework with a fourth-order advection scheme for momentum, the QUICK scheme for scalars, and second-order centered differences for computing SGS terms. In contrast, \cite{mellado18} use a DNS code with finite-difference approximations based on spectral-like sixth-order compact Padé schemes~\citep{lele92}. The latter work makes use of a significantly higher resolution due to the use of DNS \citep{pope2000, duran19}. The different codes features can influence the trend of error reduction, as the figure shows. A possible explanation for the discrepancy with the results by \cite{matheou14, matheou19} is the combination of the AMD SGS model and the pseudo-spectral advection scheme adopted in this work.

\begin{figure}[t]
\centering
\includegraphics[width=\textwidth]{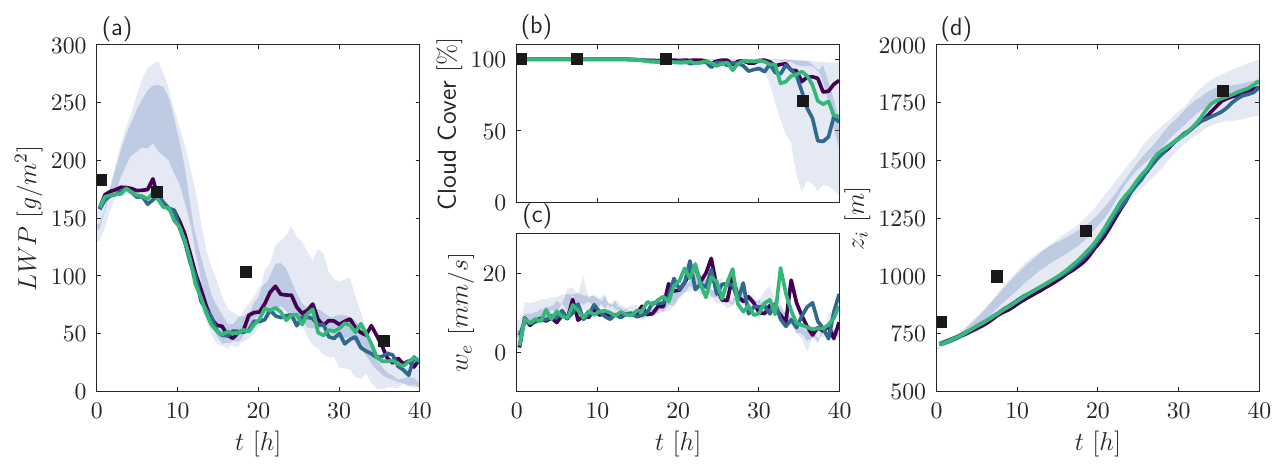}
\includegraphics[width=0.75\textwidth]{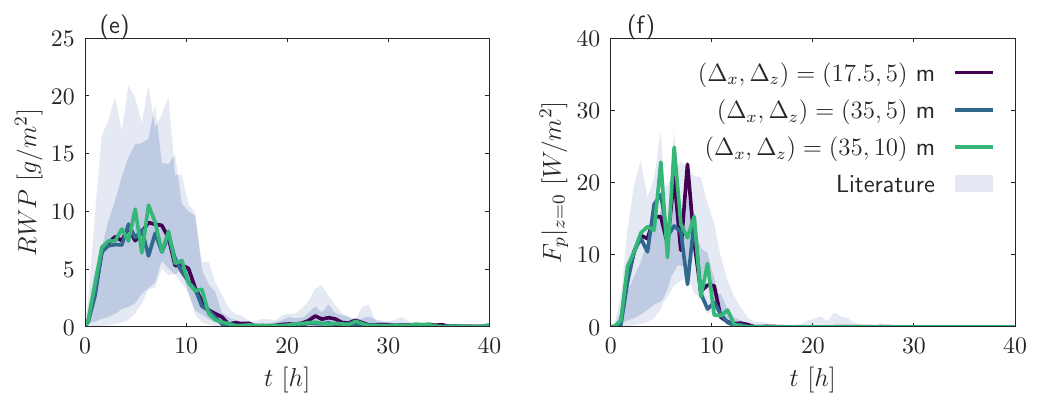}
\caption{Effect of grid resolution in the ASTEX transition case. (a): $\lwp$, (b): 30 min bin-averaged cloud cover, defined as the fraction of points in which $\lwp > \SI{13}{g/m^2}$, (c): 30 min bin-averaged entrainment velocity, (d): cloud-top height, (e): 30 min bin-averaged rain water path, (f): 30 min bin-averaged surface precipitation flux. Light shading is delimited by the maximum and minimum value within the results of different codes reported by \cite{dussen13}; dark shading denotes the central half of the distribution and is delimited by the first and third quartiles.}
\label{fig:astex_tot}
\end{figure}

\begin{figure}[!ht]
\centering
\includegraphics[width=\textwidth]{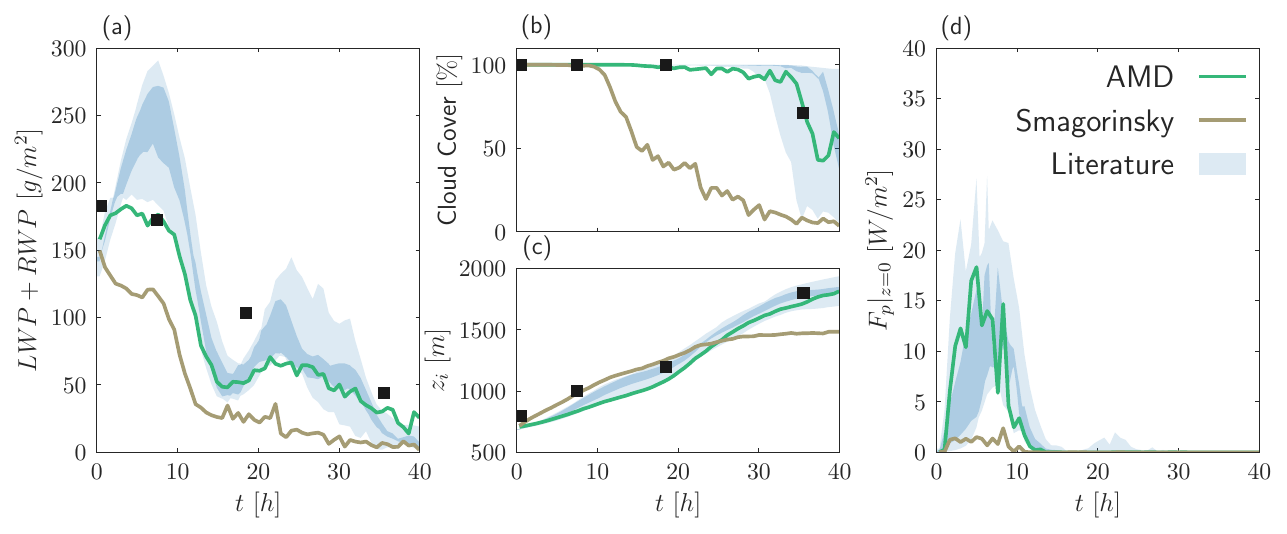}
\caption{Comparison between the AMD and Smagorinsky model simulations for the ASTEX scenario. Simulations are conducted on a uniform grid with horizontal and vertical resolutions of $(\Delta_{x,y}, \Delta_{z}) = (35, 10)$ m. (a): combined $\lwp$ and rain water path, (b): cloud-top height, (c): cloud cover, and (d) 30 min bin-averaged surface precipitation flux. For details on observational data, markers, and shaded regions, refer to the caption of Figure~\ref{fig:astex_tot}.}
\label{fig:sgs_astex}
\end{figure}

\subsection{Stratocumulus to cumulus transition: \textit{ASTEX}} \label{sec:astex}
The ASTEX setup involves a 40-hour simulation in which subsidence, sea-surface temperature, and geostrophic winds evolve dynamically over time. The initial conditions considered for this test case are reported in Appendix~\ref{app:astex} and in previous works \citep{dussen13,dussen15-thesis,albrecht95,breth99b,matheou11}. Following \cite{dussen13} and \cite{sandu11} we set the domain size to $L_x = L_y = \SI{4.48}{km}$. The vertical length is $L_z = \SI{3.2}{km}$ with a uniform vertical grid spacing up to \SI{2}{km} and increasing gradually above this height.

For this test case we consider three grid resolutions, namely $256 \times 256 \times 464$, $128 \times 128 \times 464$, and $128 \times 128 \times 256$. The horizontal grid spacing in the second case corresponds to the one adopted by \cite{dussen13}. In the uniform region, $\Delta z = \SI{5}{m}$ for the two finest simulations, and $\Delta z = \SI{10}{m}$ for the coarsest. A Rayleigh damping is used above $z = \SI{2.8}{km}$ and following \cite{breth99b}, a relaxation factor to balance the effect of subsidence warming is applied in the free-troposphere. The three simulations have average time steps of \SI{0.22}{s}, \SI{0.12}{s}, and \SI{0.11}{s}, respectively; the similarity between the latter two values reflects the CFL condition in the vertical direction to be most restrictive in determining the time step, as discussed in Section~\ref{sec:dycoms}. 

\subsubsection{Effect of grid resolution and grid anisotropy} 
The temporal evolution is shown in Figure~\ref{fig:astex_tot}, where results from the present study are compared with those from various LES codes, as reported by \cite{dussen13}. In our simulations, during the first seven hours of the simulation, the $\lwp$ increases from an initial value of approximately \SI{140}{g/m^2} to a maximum of \SI{177}{g/m^2}, closely resembling the measurement data points and the results obtained using DALES \citep{dussen13}. In this time frame, the different LES models of the dataset show a wide spread in $\lwp$ values, with maximum values reaching up to \SI{290}{g/m^2}. This variability can be attributed to differences in the rain microphysics scheme adopted \citep{dussen13}.

Figure~\ref{fig:astex_tot} also shows the rainwater path $\text{RWP} = \int_0^{L_z} \rref q_r dz$ and the surface precipitation flux $F_p\big|_{z=0} = L_v \upsilon{\phi} q_r \big|_{z=0}$. Both quantities peak during the early hours, contributing to the subsequent reduction in $\lwp$ (see Section~\ref{sec:gov_eq} and \cite{heus10}). Our results agree well with the available data from the other simulations and measurements for both cloud and boundary-layer evolution (panels~a-d), and for the precipitation (panels~e and f). Consistent with Section~\ref{sec:aniso}, the results remains reasonably accurate even on coarse meshes, demonstrating the robustness of the numerical framework.

\subsubsection{Effect of SGS model}
To evaluate the role of the SGS model in this dynamically evolving boundary layer with precipitation, Figure~\ref{fig:sgs_astex} compares the AMD model with the Smagorinsky model, using the same coefficient specified in Section~\ref{sec:smag}. Panel~(a) shows that, consistently with the DYCOMS-II RF01 case, the Smagorinsky model underestimates both the $\lwp$ and the rainwater path, RWP, defined similarly to the $\lwp$ (Equation \ref{eq:lwp}) but substituting $q_l$ with $q_r$. This underprediction is accompanied by a reduction in cloud cover, defined as the percentage of vertical columns with $\lwp > \SI{13}{g/m^2}$ (panel~b). The AMD simulation reproduces the observed cloud breakup near $t \approx 35$~h, whereas the Smagorinsky model predicts an earlier breakup around $t \approx 12$~h. This large discrepancy is attributed to the artificial evaporation of the cloud, whose liquid water content is monotonically decreasing over time as a consequence of excessive turbulence dissipation introduced by the Smagorinsky model. These differences also influence the velocity statistics (not shown) and the cloud-top height (panel~c): the Smagorinsky run concludes with $z_i \approx 1500$~m, while the AMD simulation reaches $z_i\approx 1820$~m. Lastly, the Smagorinsky model underestimates surface precipitation (Figure~\ref{fig:sgs_astex}d).


\section{Conclusions} \label{sec:conclusions}
In this work we detail the effect of grid resolution, grid anisotropy and SGS models on the accuracy of large-eddy simulation (LES) for stratocumulus-topped boundary layers (STBLs). Specifically, two STBLs field campaigns were considered, DYCOMS-II RF01 and ASTEX~\citep{stevens05, dussen13}, representative of non-precipitating and precipitating stratocumulus, respectively.

For the analysis, we developed a LES code based on spectral discretization in the horizontal and second-order finite differences in the vertical. Pseudo-spectral LES has rarely been applied to STBLs~\citep{moeng86, stevens05}, and here we revisit this approach using the anisotropic minimum dissipation (AMD) subgrid-scale model, which is well suited to anisotropic grids and does not require any parameter tuning \citep{rozema15}. As it inherently accounts for grid anisotropy, AMD is well-suited for STBLs, in which turbulence is generated mainly within a thin cloud-top layer rather than at the surface, and in this layer thermal stratification and radiative cooling exhibit sharp vertical gradients.

By defining two error metrics based on the available measurements of $\lwp$ and $\aver{w^2}$, we find that, for isotropic grid spacings, the error scalings are consistent with those obtained on numerical and theoretical grounds by \cite{duran19} for LES of turbulent planar channel flow. In particular, over the broad parameter space explored in the DYCOMS-II RF01 case, both $e_{||w||_2}$ and $e_{||lwp||_1}$ follow a scaling of the type $\Delta_\mathrm{eq}^\alpha$ with $\alpha \approx 3/4$. When anisotropic grids are used, the error reduction can be substantially faster, with scalings reaching $\alpha \approx 3$ for both error metrics. We further find that a horizontal-to-vertical grid spacing ratio between 2 and 4 yields a favorable balance between computational efficiency and accuracy. Finally, the results indicate that vertical grid spacings less than $\Delta_z \sim 10$ m are found to be sufficient for accurate results. 

In conclusion, the AMD model combined with anisotropic grids proves essential for achieving accurate and robust results across resolutions, demonstrating the value of this numerical framework for STBL simulations. This study shows that, while high resolution remains necessary to capture the small-scale features of this complex system, the choice of numerical framework is equally critical when both accuracy and computational efficiency are sought.

\section{Acknowledgements}
This project has received funding from the European Research Council (ERC) under the Horizon Europe program (Grant No.~101124815). We acknowledge the EuroHPC Joint Undertaking for awarding the project EHPC-REG-2023R03-178 access to the EuroHPC supercomputer Discoverer, hosted by Sofia Tech Park (Bulgaria). We acknowledge the Gauss Centre for Supercomputing e.V. for project ID pr74sa on the GCS Supercomputer SuperMUC-NG at the Leibniz Supercomputing Centre. We acknowledge the Dutch national e-infrastructure of SURFsara, a subsidiary of SURF, the collaborative ICT organization for Dutch education and research. We acknowledge the project `2024.056’ of the research programme “Computing Time on National Computing Facilities”, co-funded by the Dutch Research Council (NWO).

\section*{Appendix 1: Initial conditions for DYCOMS-II RF01}
\label{app:dycoms}
The geostrophic wind components are $u_G = \SI{7}{m/s}$ and $v_G = \SI{-5.5}{m/s}$, the sea-surface temperature is $\SI{292.5}{K}$, the surface pressure is $\SI{1017.8}{hPa}$. The subsidence rate is constant at $D = \SI{3.75e-6}{s^{-1}}$, and the initial boundary-layer height is $z_i = \SI{840}{m}$. The Coriolis parameter is set to $f_C = \SI{8e-5}{rad/s}$, and the surface roughness length $z_0$ is estimated using Charnock's parameterization \citep{charnock55}. Longwave radiative forcing is represented by Equation~\eqref{eq:rad}, with parameters $F_0 = \SI{70}{W/m^2}$, $F_1 = \SI{22}{W/m^2}$, and $k = \SI{85}{m^2/kg}$ \citep{stevens05}.

\begin{figure}[t!]
\centering
\includegraphics[width=\textwidth]{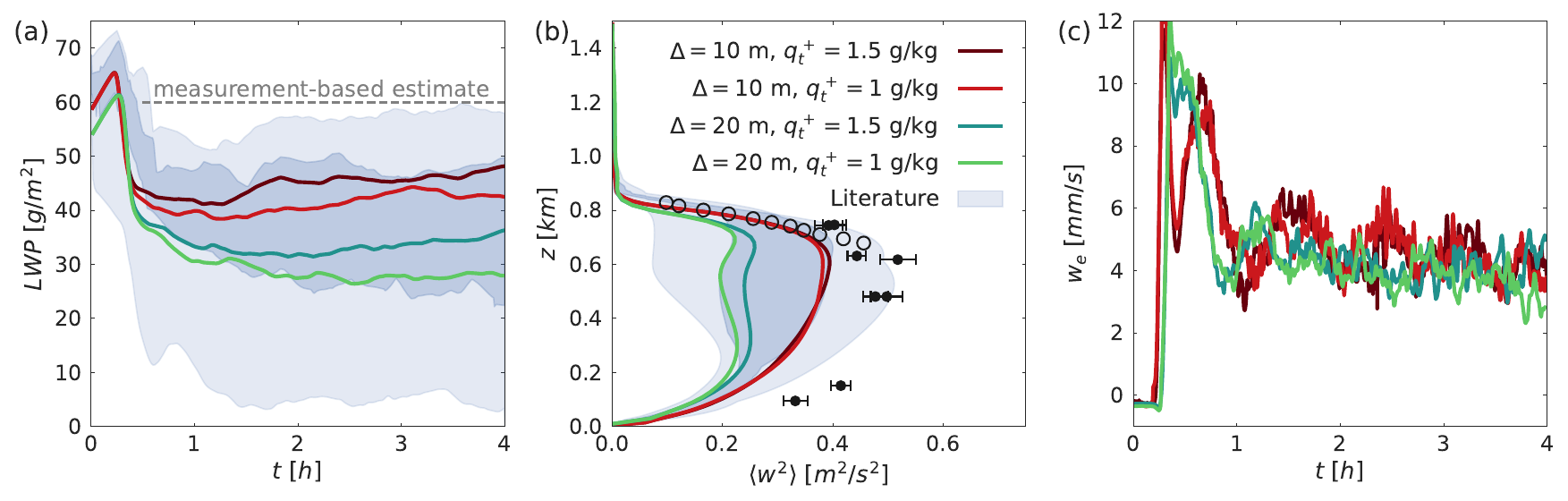}
\caption{Comparison of (a) $\lwp$, (b) vertical velocity variance, and (c) entrainment velocity obtained using $q_t=\SI{1}{g/kg}$ and $q_t=\SI{1.5}{g/kg}$ in the DYCOMS-II RF01 case. Please refer to the caption of Figure~\ref{fig:res_lwp_vel} for details on measurements, markers, and shadings.}
\label{fig:qt}
\end{figure}

Below $z_i$, the initial conditions are set to $q_t = \SI{9}{g/kg}$ and $\theta_l = \SI{298}{K}$. For $z > z_i$, the profiles follow $\theta_l = \SI{298}{K} + (z - z_i)^{1/3}$ and $q_t = \SI{1}{g/kg}$. We note that \cite{stevens05} used a value of \SI{1.5}{g/kg}. The value adopted here falls within the reported uncertainty range (see \cite{stevens03}, p.3477) and is more consistent with observational data (Figure \ref{fig:res_anel}). Figure~\ref{fig:qt} shows the effect of using two different tropospheric values for total water specific humidity: $\SI{1}{g/kg}$ used in this study, and $\SI{1.5}{g/kg}$ adopted by \cite{stevens05}. A lower free-tropospheric moisture content above the boundary layer slightly reduces $\lwp$ (panel~a) and the vertical velocity variance $\langle w^2 \rangle$ (panel~b). This occurs because moist air entrained from the free troposphere alters the local water content within the boundary layer. However, the impact remains small overall. This is also reflected in the similar entrainment velocities shown in panel~(c) \citep{stevens03}.

\section*{Appendix 2: Effect of gradient-steepness in the initial conditions of $q_t$ and $h$} \label{app:smooth}

\begin{figure}[t]
\centering
\includegraphics[width=\textwidth]{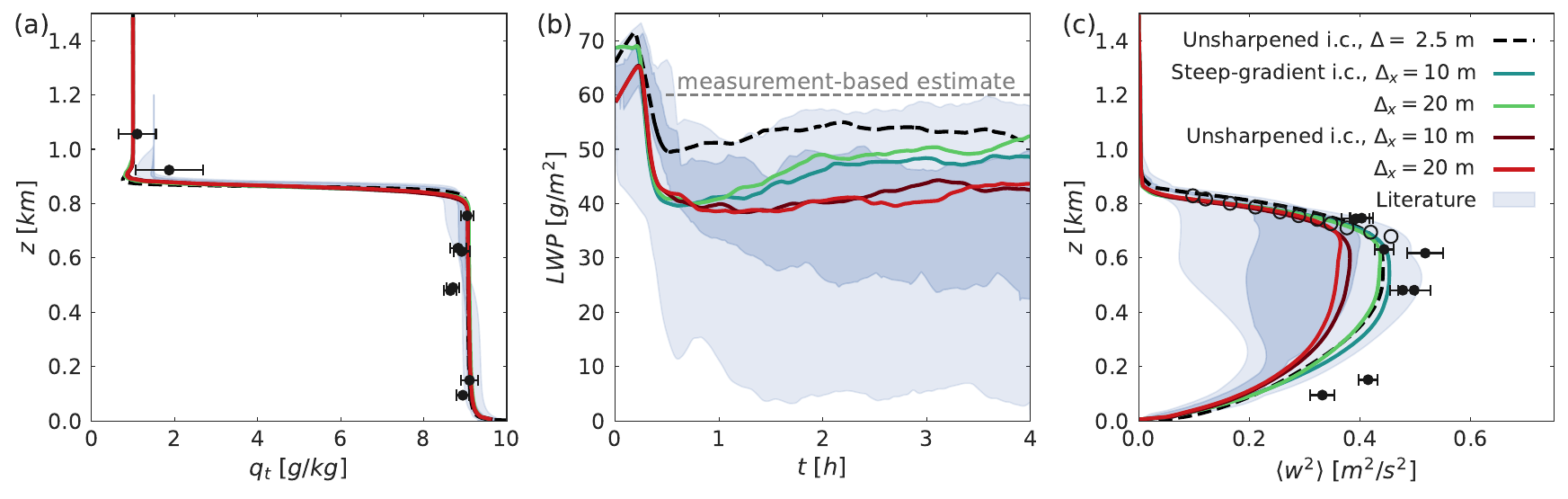}
\caption{Effect of smoothing the initial conditions in the DYCOMS-II RF01 case. Simulations employ $\Delta_z = \SI{10}{m}$, except for the reference case with $\Delta=\SI{2.5}{m}$. (a): total water humidity, (b): $\lwp$, (c): vertical velocity variance. Results in panels (a), (c) are averaged over the fourth hour. Refer to the caption of Figure~\ref{fig:res_lwp_vel} for details on measurements, markers, and shadings.}
\label{fig:smooth}
\end{figure}

In this test case we smooth the initial profiles of $q_t$ and $h$ at the inversion-layer height $z_i$ as often done in numerical simulations in which sharp gradients in the initial conditions are present. Specifically, the inversion layer of the scalars $q_t$ and $h$ is initialized using a quadratic interpolation at the inversion height. This has a stronger effect on the coarsest simulations, and lower in the finest ones, enabling us to better visualize the grid dependency of the results and to have a more controlled investigation of anisotropic grids. Other techniques can be used to improve the resolution of the gradients in the inversion-layer region, e.g. employing a linear increase of the initial conditions within a few tens of meters, as done in the ASTEX test case (Appendix~\ref{app:astex}; \cite{dussen13}).

Figure~\ref{fig:smooth} illustrates its impact on the evolution of $\lwp$ and $\langle w^2 \rangle$, using the highest-resolution case (isotropic grid with $\Delta = \SI{2.5}{m}$ spacing) as reference. While the $q_t$ profiles show only minor differences (panel~a), panel~(b) demonstrates that $\lwp$ increases substantially when initial profiles are left unsmoothed (referred to as \textit{steep-gradient i.c.} in Figure~\ref{fig:smooth}), by almost 18\% for the anisotropic case with $(\Delta_{x,y}, \Delta_z) = (\SI{20}{m}, \SI{10}{m})$ and by 11\% for the isotropic $\Delta = \SI{10}{m}$ case, when averaged over the last hour. Similarly, panel~(c) shows that $\langle w^2 \rangle$ increases and aligns with observations even at $\Delta = \SI{10}{m}$, with $\int_0^{L_z} \langle w^2 \rangle\, dz$ increasing by 16\% and 17\% for the anisotropic and isotropic cases, respectively. The magnitude of this sensitivity depends on the degree of smoothing but becomes particularly pronounced for coarser grids; for example, with $\Delta = \SI{20}{m}$, $\lwp$ increases by 40\% and the integrated velocity variance by 30\% when the profiles are left unsmoothed (not shown).

\section*{Appendix 3: Computational details for DYCOMS-II RF01} \label{app:sim_list}
Table \ref{tab:lwp_w2} reports all simulations presented in Section~\ref{sec:dycoms}, along with the main quantitative statistics and simulation characteristics. A graphical representation of the time step and computational time (last two columns of the table) with respect to grid resolution is shown in Figure~\ref{fig:scaling}.

\begin{table}
\centering
\begin{tabular}{l|c|c|c|c|c|c}
\toprule
Grid structure & AR & SGS model & \makecell{$\lwp$ \\ $[\si{g/m^2}]$} & \makecell{$ \int_{0}^{L_z} \aver{w^2} \ dz$ \\ $[\si{m^3/s^2}]$} & {$\Delta t$ [s]} & $\displaystyle \log \left ( \frac{T}{T_\mathrm{max}} \right )$ \\
\midrule
$\Delta = \SI{2.5}{m}$ & 1 & AMD & 52.97 & 288.5 & 0.033 & 1 \\
\midrule
$\Delta = \SI{5}{m}$ & 1 & AMD & 50.00 & 273.2 & 0.066 & $5.6 \times 10^{-2}$ \\
\midrule
$\Delta_{x,y} = \SI{10}{m}$, $\Delta_z = \SI{2.5}{m}$ & 4 & AMD & 52.28 & 296.7 & 0.070 & $1.4 \times 10^{-2}$ \\
\hspace{2.14cm} $\Delta_z = \SI{5.0}{m}$ & 2 & AMD & 51.26 & 279.4 & 0.131 & $3.4 \times 10^{-3}$ \\
\hspace{2.14cm} $\Delta_z = \SI{10}{m}$ & 1 & AMD & 43.19 & 245.9 & 0.135 & $1.6 \times 10^{-3}$ \\
\midrule
$\Delta_{x,y} = \SI{20}{m}$, $\Delta_z = \SI{2.5}{m}$ & 8 & AMD & 48.10 & 254.6 & 0.079 & $3.1 \times 10^{-3}$ \\
\hspace{2.14cm} $\Delta_z = \SI{5.0}{m}$ & 4 & AMD & 47.39 & 247.9 & 0.158 & $6.7 \times 10^{-4}$ \\
\hspace{2.14cm} $\Delta_z = \SI{10}{m}$ & 2 & AMD & 42.63 & 238.6 & 0.272 & $1.8 \times 10^{-4}$ \\
\hspace{2.14cm} $\Delta_z = \SI{20}{m}$ & 1 & AMD & 27.88 & 155.8 & 0.275 & $8.4 \times 10^{-5}$ \\
\midrule
$\Delta = \SI{10}{m}$ & 1 & Smagorinsky & 12.30 & 119.2 & 0.138 & $1.5 \times 10^{-3}$ \\
\midrule
\makecell{$\Delta_{x,y} = \SI{20}{m}$, $\Delta_z = \SI{10}{m}$}
& 2 & Smagorinsky & 8.27 & 108.9 & 0.279 & $1.6 \times 10^{-3}$ \\
\midrule
\makecell{Steep-gradient i.c., \\ \hspace{2.3cm} $\Delta = \SI{10}{m}$} & \makecell{\\1} & \makecell{\\AMD} & \makecell{\\48.04} & \makecell{\\288.0} & \makecell{\\0.135} & \makecell{\\$1.01 \times 10^{-1}$} \\
\midrule
\makecell{Steep-gradient i.c. \\ $\Delta_{x,y} = \SI{20}{m}$, $\Delta_z = \SI{10}{m}$} 
& \makecell{\\2} & \makecell{\\AMD} & \makecell{\\50.27} & \makecell{\\277.3} & \makecell{\\0.271} & \makecell{\\$1.8 \times 10^{-4}$} \\
\hspace{2.2cm} $\Delta_z = \SI{20}{m}$ & 1 & AMD & 39.17 & 203.3 & 0.274 & $8.4 \times 10^{-5}$ \\
\bottomrule
\end{tabular}

\caption{Description of the simulations for the DYCOMS-II RF01 case. Columns from left to right: simulations performed, aspect ratio $\Delta_x/\Delta_z$, SGS model adopted, liquid water path, integral of vertical velocity variance, averaged time step, CPU-hours per simulation normalized by the total resolution $N$.}
\label{tab:lwp_w2}
\end{table}

\section*{Appendix 4: Initial conditions for the ASTEX case}
\label{app:astex}
The initial conditions of $q_t$ and $h$ are defined according to the ones described by \cite{dussen13}. The initial cloud-top height is $z_i = \SI{662.5}{m}$, surface pressure is fixed at \SI{1029}{hPa}, the Coriolis parameter is set to \SI{8.7e-5}{rad/s}, and the surface roughness length is $z_0 = \SI{2e-4}{m}$. Below the initial $z_i$, we set $q_t=\SI{10.2}{g/kg}$ and $\theta_l=\SI{288}{K}$; for $z_i<z<z_i+50$~m, a difference of $\Delta q_t=\SI{-1.1}{g/kg}$ and $\Delta \theta_l=\SI{5.5}{K}$ is applied, after which a constant lapse rate of $\Gamma_{q_t}=\SI{-2.8}{g/kg/km}$ and $\Gamma_{\theta_l}=\SI{6}{K/km}$ is set up to the domain top \citep{dussen15}. Longwave radiative cooling is modeled using Equation~\eqref{eq:rad} with $F_0 = \SI{74}{W/m^2}$, $F_1 = \SI{15}{W/m^2}$, and $\kappa = \SI{130}{m^2/kg}$ \citep{chung12}. Shortwave radiative forcing is modeled using a solar constant $S_0=\SI{1376}{W/m^2}$. The cloud droplet number concentration is set to $N_c = \SI{100}{cm^{-3}}$ and is used to model cloud droplet sedimentation and precipitation \citep{dussen13}. Precipitation is parameterized as described in Section~\ref{sec:gov_eq}. The time step is constrained by the fixed CFL, which is chosen to ensure that the maximum distance traveled by drizzle within a time step does not exceed one vertical grid length. 

\bibliographystyle{plainnat}
\bibliography{bib.bib}

\end{document}